\documentclass[ aps,%
floatfix,%
final,%
notitlepage,%
oneside,%
onecolumn,%
nobibnotes,%
nofootinbib,%
superscriptaddress,%
showpacs,%
]%
{revtex4}

\newcommand{\G}{\Gamma} 
\newcommand{\R}{\mathcal{R}}
\newcommand{\A}{\mathcal{A}}
\newcommand{\M}{\mathcal{M}}

\newcommand{\qP}{q_\Vert} \newcommand{\qpP}{q'_\Vert}
\newcommand{\qT}{\mathbf{q}_\perp} \newcommand{\qpT}{\mathbf{q}'_\perp}

\newcommand{\JP}{\psi} \newcommand{\ec}{\eta_c}
\newcommand{\epem}{e^+e^-}

\newcommand{\MeV}{\,\mathrm{MeV}}
\newcommand{\GeV}{\,\mathrm{GeV}}
\newcommand{\fb}{\,\mathrm{fb}}\newcommand{\nb}{\,\mathrm{nb}}\newcommand{\mcb}{\,\mu\mathrm{b}}
\newcommand{\pb}{\,\mathrm{pb}}

\newcommand{\Br}{\mathrm{Br}}

\newcommand{\beq}{\begin{eqnarray}}\newcommand{\eeq}{\end{eqnarray}}
\newcommand{\beqa}{\begin{eqnarray*}}\newcommand{\eeqa}{\end{eqnarray*}}

\usepackage{graphicx}
\usepackage{epsfig}

\begin{document}

\title{Observation potential for $\chi_b$ at the Tevatron and LHC}
\author{V.V. Braguta}
\email{braguta@mail.ru}

\author{A.K. Likhoded}
\email{Likhoded@ihep.ru}

\author{A.V. Luchinsky}
\email{Alexey.Luchinsky@ihep.ru}
\affiliation{Institute for High Energy Physics, Protvino, Russia}

\begin{abstract}
We confirm the results of previous works that the internal motion of quarks inside charmonium
mesons increases the cross
section of the process $\epem\to J/\psi\eta_c$. We also show, that this effect increases the
widths of the scalar meson
decay into two vector ones and state that the decays $\chi_{b0,2}\to2J/\psi$ can be used to
detect these scalar mesons at
Tevatron and LHC colliders.
\end{abstract}

\pacs{
12.38.-t,  % Quantum chromodynamics ... ... Quarks, gluons, and QCD in nuclei and nuclear
% processes
12.38.Bx,  % Perturbative calculations
13.66.Bc,  % Hadron production in e-e+ interactions
13.25.Gv % Decays of J/psi, Upsilon, and other quarkonia
}

\maketitle

%%%%%%%%%%%%%%%%%%%%%%%%%%%%%%%%%%%%%%%%%%%%%%%
\section{Introduction}

Recently Belle collaboration has studied the production of pseudoscalar and vector charmonium
mesons in electron-positron
annihilation at $\sqrt{s}=10.6\GeV$ \cite{Abe:2002rb}. The lower bound of the cross section
measured at Belle
\beqa
  \sigma(\epem\to\JP\ec) & > & 33\fb
\eeqa
is about an order of magnitude higher, than the theoretical predictions \cite{Braaten:2002fi}.
Some efforts were made to
explain this discrepancy. For example, in \cite{Bodwin:2002fk} it was assumed, that some of the
Belle's $\JP\ec$ signal
could actually be $\JP\JP$ events and the value of the cross-section of two photon process
$\epem\to\JP\JP$ was presented.
However in later works \cite{Bodwin:2002kk,Luchinsky:2003yh} it was shown that QCD corrections
decrease this cross section
and the subsequent experimental \cite{Abe:2003ja} analysis excluded this possibility completely.
Another possible
exclamations could be the contributions of glueball \cite{Lee:2003db} and color octet states to
this process

Lately the discrepancy between theoretical predictions and experimental value for the cross
section of the process
$\epem\to\JP\ec$ has found the surprisingly simple explanation. In the works \cite{
Ma:2004qf,Bondar:2004sv}  it was
shown, that taking into consideration intrinsic motion of quarks inside mesons one can
significantly decrease the
virtuality of intermediate particles and hence raise the value of the cross section. The method
of light cone expansion,
used in these works, is actually an expansion in a small parameter $r\sim M / \sqrt s$, where $M$
is the meson mass. It
should be mentioned however that this method can lead to large uncertainties, caused by poor
knowledge of quark
distribution functions. As it was shown in \cite{Ma:2004qf} one can significantly change the
result by changing the
expressions of these functions. Another method, used prior to these works was the expansion in
internal velocity $v$ of
quarks in a meson. In most of the works considering the process $\epem\to\JP\ec$ only the leading
term of this expansion
was used and the decrease of virtuality of internal particles was neglected. The purpose of our
paper is to consider of
this effect in the framework of amplitude expansion over the internal velocity $v$ and comparison
of our results with light
cone approximations.

Besides the reaction $\epem\to\JP\ec$ there are also other processes in which the intermediate
particles propagate with
large virtuality when one neglects the internal motion of quarks inside mesons. Among these
processes  we would like to
mention the decays of scalar and tensor mesons into two vector ones, i.e. $\chi_{0,2}\to VV$. In
this paper we will
consider some of these decays, namely $\chi_{c0,2}\to\rho\rho,\,\phi\phi$ and
$\chi_{b0,2}\to\JP\JP$. The main reasons for this interest are:
\begin{itemize}
\item
The only known experimental
value for such processes \cite{Bai:1998cw}
\beq
  \Br\left(\chi_{c0}\to\phi\phi\right) & = & \left(1.0\pm0.6\right)\cdot10^{-3} \label{eq:Br}
\eeq
is much greater than the theoretical predictions made in the assumption that the intrinsic motion
of quarks can be neglected (in what follows this assumption will be called
"$\delta$-approximation")
\cite{Kartvelishvili:1984en,Zhou:2004mw} and one could expect that taking into account internal
motion would increase them.

\item

As it was mentioned in \cite{Kartvelishvili:1984en}, the decays of $\chi_{b0,2}$ can  be used to
study these mesons in
high energy experiments (for example in the reaction $p\bar p\to\chi_{bJ}X\to\JP\JP X$). The
branching ratio of the decay
$\chi_{b0}\to\JP\JP$ obtained in this work
\beqa
\Br(\chi_{b0}\to\JP\JP) & =  &3\cdot10^{-5}
\eeqa
is however too small to use this mode in studying the properties of $\chi_{b0}$ meson. When
estimating this branching
fraction the authors have neglected the internal motion of $c$ quarks in the hard part of the
amplitude. As it will be
shown bellow, taking it into account one can increase the value of this branching
fraction by an order of
magnitude. In this conditions the observation of $\chi_b$ mesons in $\JP\JP$ mode is a feasible
task.
\end{itemize}

%%%%%%%%%%%%%%%%%%%%%%%%%%%%%%%%%%%%%%%%%%%%%%%%%%%%%%%%%%%%%%%%%%%
\section{$\protect\epem\to\JP\ec$}

\begin{figure}
  \begin{picture}(400,560)
  \put(0,0){\epsfxsize=13cm \epsfbox{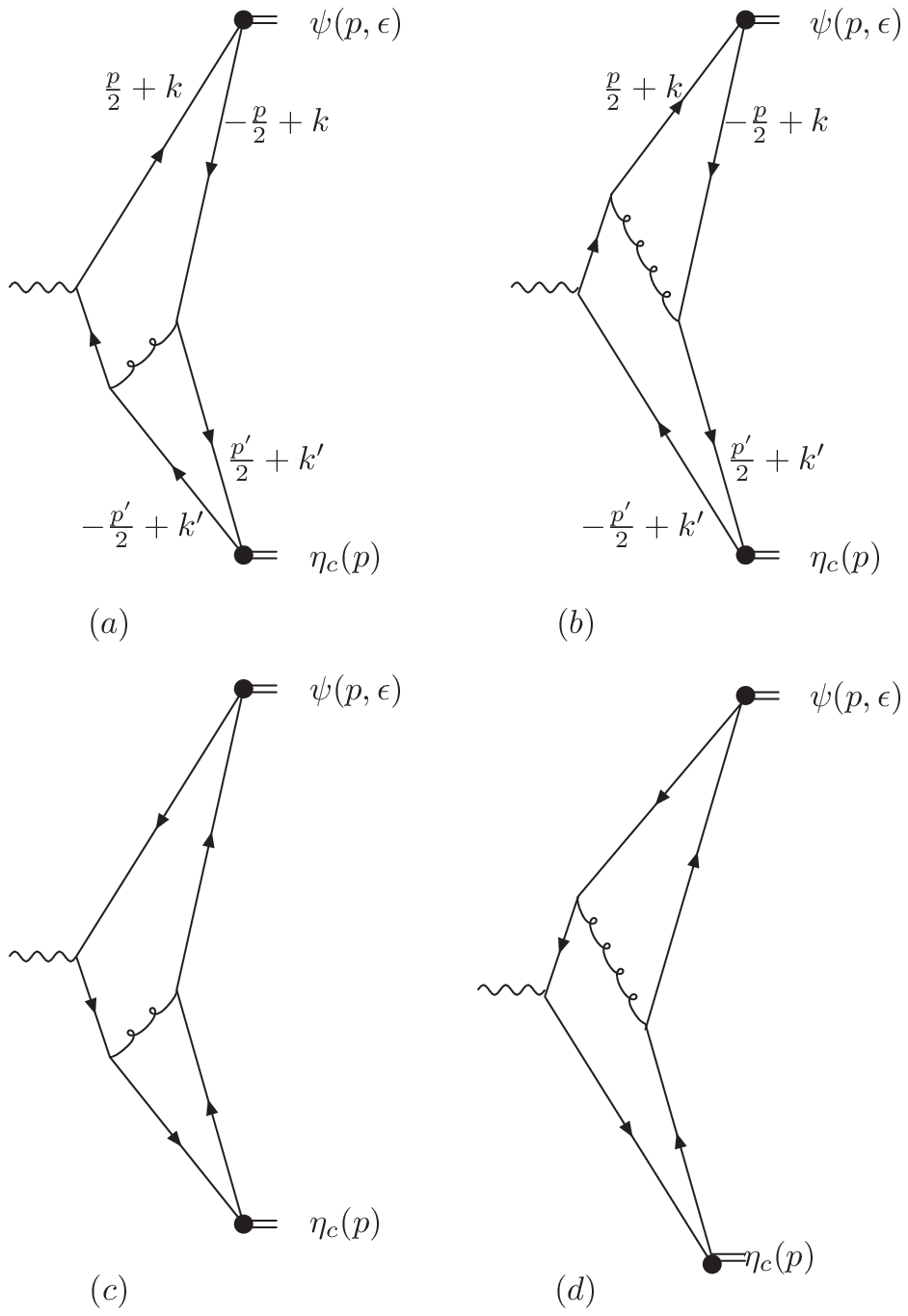}}
  \end{picture}
  \caption{\protect$\epem\to\JP\ec$}
  \label{fig:eePsiEta}
\end{figure}

The diagrams contributing to the process $\epem\to\JP\ec$ at the leading order in the strong
coupling constant $\alpha_s$
are shown on fig.\ref{fig:eePsiEta} and the corresponding amplitude can be written in the form
\cite{Braaten:2002fi}:
\beqa
  \M & = &
    - \frac { e_c e^2 } { s } \bar v(k_2) \gamma^{\mu} u(k_1)
    \left<
      \JP + \ec | J_{\mu}(0)| 0 \right>.
\eeqa
If one neglects the internal motion of quarks in mesons, the matrix element of the
electromagnetic current is equal to
\beq
  \left<\JP(p,e) + \ec(p') | J^{\mu}(0)| 0 \right> & = &
  i \frac {1024 \pi \alpha_s} { 3 s^2} \Psi_{\eta_c} (0) \Psi_{\JP} (0)
  \epsilon^{\mu \nu \lambda \sigma} p_{\nu} p'_{\lambda} e_{\sigma},
  \label{nrqcd}
\eeq
where $\Psi_{{\JP},\ec}(0)$ are the values of $\JP$ and $\ec$ wave functions in the origin. In
this approximation the value
of the cross section of this process is $\sigma = 1.9$ fb, about an order of magnitude smaller,
than the experimental
result \cite{Abe:2002rb}. As it was mentioned above, a possible reason for this discrepancy is
that internal motion of
quarks was neglected in the hard scattering part of this process. This effect was studied in the
framework of light cone
expansion \cite{Ma:2004qf,Bondar:2004sv}, i.e. an expansion in the parameter
$r\sim M / \sqrt {s} \sim 1/3$. The
calculations show, that one increases greatly the value of the cross section and allows to
achieve the same values.
It should be mentioned, that the method used in these works can lead to uncertainties caused by
the poor knowledge of the
distribution functions. One can obtain almost arbitrary result by changing these functions \cite{
Ma:2004qf}.

Another method used in studying the process $\epem\to\JP\ec$ is the expansion of the amplitude in
the relative velocity $v$
of quarks in a meson. The potential models \cite{Buchmuller:1980su} give the value $v^2=0.23$, so
the expansion
parameter $v \sim 0.5$, that is certainly larger, than the expansion parameter of the light cone
expansion formalism.
In our article we intend to consider the decrease of virtuality of intermediate quarks and gluons
in the framework of the
following simple model. The contribution of the diagram fig.\ref{fig:eePsiEta}a to the amplitude
of the $\epem$
annihilation into two $c\bar c$ pairs and the hadronization of these pairs into final mesons
$\JP$ and $\ec$ is
proportional to
\beq
  \M & \sim &  l^{\mu}
  \frac {
    \bar u ( p'/2 + k')\gamma^{\nu} v (p/2 - k)
  }{
    ( (p+p')/2 + k' - k )^2 (( p + p'/2 + k' )^2-m_c^2)
  }
  \times\nonumber \\
  & \times &
  \bar u\left(\frac p 2 +k\right) \gamma_{\mu}
  \left(\hat k - \hat {p'} -\frac { \hat p}{2} +m_c \right)
  \gamma_{\nu} v\left( \frac{p'} 2 - k'\right),
\label{k}
\eeq
where $k$ and $k'$ are the relative momenta of quarks in $\JP$ and $\ec$ mesons, $u$ and $v$ are
quark and antiquark
bispinors and $l^\mu$ is the leptonic current. If the motion of quark-antiquark pair in the meson
is considered as
non-relativistic, one can expand the expression (\ref{k}) over the relative momenta $k$ and $k'$.
The expression
(\ref{nrqcd}) for example is derived from the first term of this expansion. Taking the higher
terms of this expansion  we
however encounter the difficulty caused by the strong dependence of propagators in eq. (\ref{k})
on $k$ and $k'$ (actually,
at large values of this variables the series will not converge at all). To avoid this difficulty
we will use the exact
expressions for the denominators in eq.(\ref{k}) and neglect internal motion of quarks in the
numerator of this expression.
In the c.m.s. of the $\JP\ec$ pair the following formulae are valid:
\beqa
  p &=& \gamma M\left\{1,{\mathbf 0}_\perp,\beta\right\}, \\
  p'&=& \gamma M\left\{1,{\mathbf 0}_\perp,-\beta\right\}, \\
  k &=& \left\{\gamma \qP \beta,\qT, \gamma \qP\right\}, \\
  k'&=& \left\{-\gamma \qpP \beta, \qpT, \gamma \qpP\right\},
\eeqa
where $M$ is the mass of the charmonium meson (for simplicity we neglect the difference between
vector and pseudoscalar
meson masses), $\beta$ is the velocity of one of the charmonia (directed along the Oz axes),
$\mathbf{q}$ and $\mathbf{q}'$
are the space parts of the relative momenta $k$ and $k'$ in the rest frames of the corresponding
mesons,  $\qP$ and $\qpP$
are z-projections of these vectors and $\qT$ and $\qpT$ are their two-dimensional components
perpendicular to this axes.
From these equations the following expressions for the scalar products can be obtained:
\beqa
  pk' &=& -2 \beta \gamma^2 M \qpP, \\
  p'k &=& 2 \beta \gamma^2 M \qP, \\
  kk' & = &  -\gamma^2 \qP\qpP (1+\beta^2) - \qT\qpT, \\
  k^2 &=& - \qP^2 - \qT^2, \\
  k'^2 &=& - \qpP - {\qpT}^2.
\eeqa
It is clear that the scalar products $(pk')$ and $(p'k)$ are enhanced by the factor
$\gamma= \sqrt s /2M$, in $(kk')$ there
are both enhanced and unenhanced terms and scalar products $k^2$ and $k'^2$ do not contain this
factor at all. In the light
cone expansion formalism the perpendicular components of vectors $\mathbf{q}$ and $\mathbf{q}'$
(i.e. terms that are not
enhanced by $\gamma$) are neglected. This approximation is valid for reactions with high c.m.
energy, where $\gamma\gg 1$,
but can lead to sufficient errors in other cases. That is why we will carry out our calculations
using two methods --- both
neglecting contributions of this terms and leaving them nonzero and afterwards we will compare
the results.

The internal motion of quarks inside $\JP$ meson will be described by setting its quark and
antiquark momenta to $P/2+k$
and $P/2-k$ respectively. Similarly, the momenta of quark and antiquark hadronizing into $\eta_c$
will be $P'\pm k'/2$.
After this substitution the gluon propagator will have the
form
\beq
  \left(\frac {p+p'} 2 + k' - k\right)^2  & = &
  \frac{s}{4} \left(
    1 + \frac{4}{s} (p'k - pk') + \frac{4}{s} (k - k')^2
  \right)
  =\frac s 4 d,
  \label{gl}
\eeq
and quark propagators on fig.\ref{fig:eePsiEta} will be equal to
\beq
  \left(\frac{p}{2} + p' - k \right)^2 - m_c^2 &=&
  \frac{s}{2} \left(
    1 - \frac{4}{s} p'k +
    \frac{2}{s} \left(
      k^2+ \frac {M^2}{4} -m_c^2
    \right)\right)
  = \frac s 2 s_1,
\label{qu1}\\
  \left(\frac {p'}{2} + p + k' \right)^2 - m_c^2 &= &
  \frac{s}{2} \left(
    1 + \frac{4}{s} pk' + \frac{2}{s} \left(
      k'^2+ \frac{M^2}{4} -m_c^2
   \right)\right) =\frac{s}{2}s_2.
\label{qu2}
\eeq
From equations (\ref{gl}), (\ref{qu1}), (\ref{qu2}) it is clear, that in $\delta$-approximation (
i.e. neglecting internal
motion and assuming the quark mass $m_c=M/2$) the dimensionless propagators $d$ and $s_{1,2}$
will be equal to 1. Keeping
this in mind it is easy to understand that to obtain the expression for any diagram of
fig.\ref{fig:eePsiEta}
one should divide the expression (\ref{nrqcd}), where this motion was neglected, by this
dimensionless propagators and
integrate it with the proper chosen wave functions. It is useful to notice, that the diagrams \ref
{fig:eePsiEta}c can be
obtained by charge conjugation of the diagram \ref{fig:eePsiEta}a, so it is not necessary to
calculate it separately (the
same is also valid for diagrams \ref{fig:eePsiEta}b and \ref{fig:eePsiEta}d) Thus, the amplitude
of the process
$\epem\to\JP\ec$ can be written in the form
\beq
  \M &=&
  \M_0 \frac 1 {\int d \rho} \int d\rho \frac 1 {2d} \left( \frac {1} {s_1} + \frac {1} {s_2}
  \right) =
  \M_0 K
\label{result},
\eeq
where the integration is hold over the wave functions of $\JP$ and $\ec$ mesons in momentum
representation:
\beqa
  d\rho  & = &  d^3 q d^3 q' \phi_{\JP}(\mathbf{q})\phi_{\ec}(\mathbf{q}').
\eeqa
In equation (\ref{result}) $\M_0$ is the amplitude of this process with no internal motion and
the factor $K$ describes the
decrease of momenta of virtual quarks and gluon. The variation of wave function width results in
change of this factor and
therefore the variation of annihilation cross section. It is clear, that if we fix the value of
the amplitude
and tend the width of the wave function to zero (i.e. neglect the internal motion of quarks),
than the factor $K$ will tend
to 1 and the equation (\ref{result}) will reduce to (\ref{nrqcd}). It should be mentioned also,
the values $\Psi_{J/ \Psi}(0)$, $\Psi_{\eta_c}(0)$ also depend on the choice of the wave
functions ---  it
is obvious, that with increase of the width of the distribution in $\mathbf{q}$ the values of the
wave function in the
origin will grow. Thus, the variation of wave function width will lead to double effect: the
variation of the momenta of
internal particles and the change of the values of wave functions in the origin. Both these
effects are taken into account
in the equation (\ref{result}).

\section{Wave functions}

In our model we assign the momentum $P/2+k$ to $c$ quark in $\JP$. It is evident, that this quark
is not on mass shell:
its virtuality depends on the relative momentum $k$ and equals 	to $M^2/4+k^2$. The formal use of
this formula can lead to
unphysical situations, when the virtuality of $c$ quark is zero or negative. It is obvious, that
such cases would
occur only if  the motion of  $c\bar c$ pair in the meson is relativistic, whereas in the model
used
here is valid only for non-relativistic movement, when the virtuality of this quark does not
differ significantly from
$M/2$. The calculations described below will show, that even in the broad wave-function case (
$v^2=0.4$) the maximum of
the amplitude is accumulated at $-k^2 \sim 1$ GeV$^2$, that corresponds to virtuality $\sim 1.2$
GeV.

Another unphysical singularity caused by the fact that 	the virtuality of quarks and antiquarks
can be zero or negative is
that the propagators (\ref{gl}), (\ref{qu1}),(\ref{qu2}) can vanish at some values of quarks'
momenta. As a result one can
obtain almost arbitrary values for the cross section of the process under consideration by using
different distribution
functions.

In the previous works the potential quark model was often used to derive the form of the wave
function. Such a function,
obtained from Schr{\"o}dinger equation solution allows, although with a small probability, any
values of the relative
momentum of quarks and we will get an infinitive result when substituting it in the equation (
\ref{result}).
This problem is obviously caused by the singularity of propagators in the region $q \sim M/2$,
where the motion of the
quarks is relativistic and the usage of non-relativistic wave functions is not valid. To avoid
this difficulty, when
deriving the expression for the wave function we will use another method, proposed in \cite{
Kartvelishvili:1985ac}.

In what follows we will restrict ourself to the valence quark approximation, i.e. we will suppose
that $\JP$ and $\ec$
mesons are built only from $c\bar c$ pair. Let the momentum of the quark in the meson rest frame
be $q=(q_0 , {\bf q} )$
and the meson's wave function in the momentum representation is $\phi (\mathbf{q})$. Then, in the
frame of infinitive
momentum the $c$ quark distribution function will be
\beqa
  f(x) & = & \int d^3 q |\phi ( {\bf q} ) |^2 \delta \left(x- \frac { q_0+ q_z } M\right),
\eeqa
where $M$ is the meson mass and $x$ is the $c$-quark momentum fraction. For the mesons with
hidden flavour in valence quark
approximation we have $q_0=M/2$. It is useful to introduce the following integration variables:
\beqa
  d^3 q & = & \frac {\pi M^2} 2 v dv d\qP,
\eeqa
where $v=2 |{\bf q}|/ M$. After integration over $q_z$
\beqa
  f(x) & = &
    \frac {\pi M^3} 4 \int_{\Phi(x)}^1 d v^2 |\phi ( v^2 )|^2,
  \\ \Phi(x) & = & (1-2x)^2.
\eeqa
Differentiating this relation with respect to $x$ we get
\beqa
  f'(x) & = & - \frac{\pi M^3}{4} \Phi'(x) |\phi(\Phi(x))|^2.
\eeqa
From the last formula one can obtain the relation between meson wave function and the structure
function in the infinitive
momentum frame:
\beq
  |\phi(v^2)|^2 & = & -\frac 4 {\pi M^3} \frac {d f(\Phi^{-1}(v^2))}{d v^2}.
\label{func}
\eeq
Let us use the Regge parametrization of the structure function \cite{
Chliapnikov:1977fc,Kartvelishvili:1977pi,Kartvelishvili:1979aa}:
\beqa
  f(x) &\sim & x^{- \alpha} (1-x)^{\gamma - \alpha}.
\eeqa
Since we restrict ourself to two-body state $c\bar c$ and neglect the contribution of the sea
quarks the parameter $\gamma$
should be set equal to zero. Substituting the expression for the wave function into (\ref{func})
we get
\beq
  \phi(q)  &=& \left(
    \frac {- \alpha} {\pi M^3 B(1-\alpha, 1-\alpha ) }
  \right)^{1/2} \left( \frac {1- v^2} 4 \right)^{(- \alpha-1)/2}.
  \label{vf1}
\eeq

Let us determine the value of the parameter $\alpha$. We can do this by fixing either the
parameters of the $\JP$
regge trajectory, or mean square of the velocity of $c$-quark in meson. In what follows we will
use the latter method.
The calculations will be held using four different values of mean quark velocity :$v^2 = 0.2$,
$0.3$, $1/3$ and $0.4$.
These values correspond to $\alpha= -6.0$, $-3.5$, $-3.0$ and $-2.25$. It is interesting to
notice, that for $v^2=1/3$,
$\alpha=-3.0$ \cite{Kartvelishvili:1977pi} the wave function integrated over transvese momentum
differs from light cone
function used in \cite{Bondar:2004sv} no more than by $5\%$. We have chosen the value $v^2=1/3$
because the calculation of
$\JP\to\epem$ and $\ec\to\gamma\gamma$ decay widths with this function leads to the values
$ \Gamma(\JP\to\epem) = 5.45$
keV, $\Gamma(\ec\to\gamma\gamma) = 7.3$ keV, that are in agreement with the experiment
\cite{Aubert:2003sv,Armstrong:1995nn}.

The results of our calculations for listed values of $v^2$ are presented in table \ref{tab:tab}.
In the first column there
are values for $|R(0)|^2 = 4 \pi |\Psi(0)|^2$, the second one contains the values of the cross
section obtained from
equation (\ref{nrqcd}) neglecting internal quark motion in the hard part of the amplitude, the
last two ones contain the
values of the factor $K^2$ from formula (\ref{result}) neglecting and leaving the transverse
components of internal momenta
$k$ and $k'$. Finally, in last column we give the values of the total cross section.

The results of our calculations, presented in table \ref{tab:tab} show, that the calculation held
with exact dependence
of propagators on internal momenta gives the values that are lager, than the values obtained
neglecting this motion. For
example, for the case $v^2=0.4$ the cross section is multiplied by 4. So our assumption that the
dependence of propagators
on the internal momenta is strong proved to be correct and the expansion of this propagators in
relative velocity $v$
leads to the series with rather poor convergence. Although, the effect described above increases
the value of the cross
section, this enhancement is not large enough to explains the experimental data obtained
at Belle. As it was mentioned above, the calculations in the framework of the light cone
expansion method \cite{
Bondar:2004sv} give the values that are in agreement with experimental data and, if there is no
other reasons that can
increase the cross section obtained in the framework of NRQCD by an order of magnitude we can
assume that the light cone
expansion parameter $M/ \sqrt{s} \sim 1/3$ is more suitable for describing the double quarkonia
states in the electron-
positron annihilation at $\sqrt{s}=10.6$ GeV, then the relative velocity $v \sim 0.5$.

Another interesting feature that can be seen from table \ref{tab:tab} is that taking into account
the terms suppressed
by $s$ leads to significant variation of the results. For the narrow wave function with $v^2=0.2$
this correction is about
$20\%$, while for the wide one with $v^2=0.4$ the cross section is almost doubled. Thus, the
transverse
components of internal momenta have to be taken into account. In the light cone expansion
formalism the terms
suppressed by $s$ are neglected and it results in a large error. However, taking this terms into
account will only increase
the one, and a new result does not contradict the experiment.

%%%%%%%%%%%%%%%%%%%%%%%%%%%%%%%%%%%%%%%%%%%%%%%%%%%%%%%%%%%%%%%%%%%
\section{$\chi_J\to VV$}

In addition to the $\epem\to\JP\ec$ process considered in the previous sections there are also
other reactions in which
intermediate particles have large virtuality when $\delta$-approximation is used. Among these
processes we would like to
note the decays $\chi_{0,2}\to VV$. In the $\delta$-approximation these decays were already
studied in literature
\cite{Kartvelishvili:1984en,Zhou:2004mw}

\begin{figure}
  \begin{picture}(400,130)
  \put(0,0){\epsfxsize=15cm \epsfbox{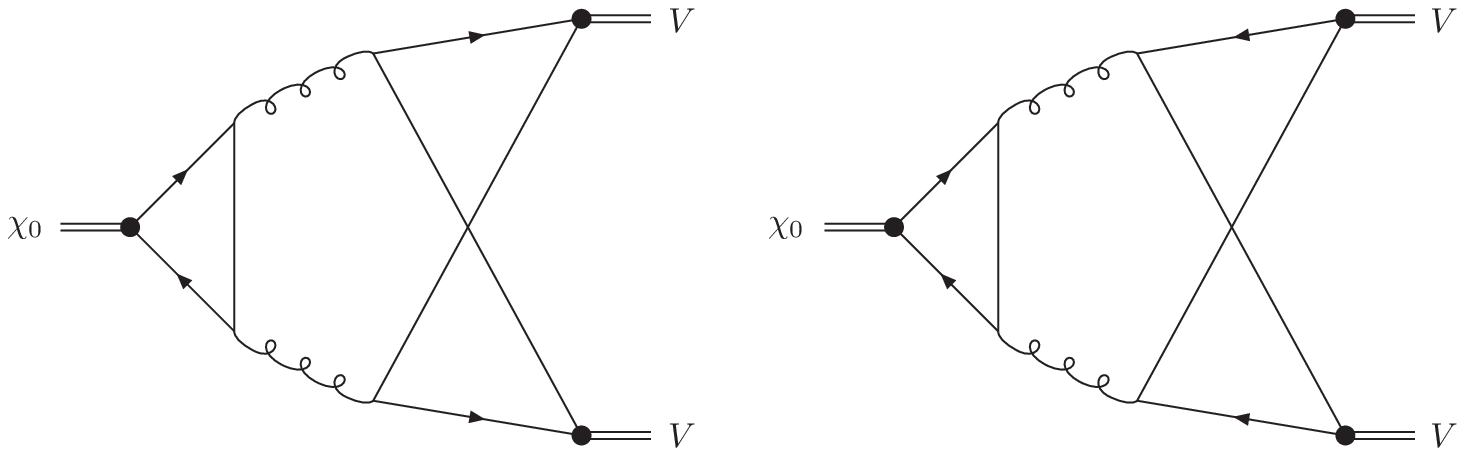}}
  \end{picture}
  \caption{\protect$\chi_J\to VV$}
  \label{fig:chiVV}
\end{figure}

Two diagrams contributing to such decays at the leading order on the strong coupling constant
$\alpha_s$ are shown
in fig.\ref{fig:chiVV}, the others can be obtained by interchanging final vector mesons. It can
be easily
seen, that the virtuality of gluons on these diagrams in the $\delta$-approximation equals
$M_\chi^2/4$ and one could
expect, that similarly to the results of previous sections it will decrease if the internal
motion of quarks in the mesons
is taken into account.

Analytical formulae for the width of the decay $\chi_0\to VV$ were presented in the work \cite{
Anselmino:1992rw} and we
will use this results in our paper. The nonzero helicity amplitudes
$\A^{(0)}_{\lambda_1,\lambda_2}$ of the decay of
scalar meson $\chi_0$ into vector mesons $V_1$ and $V_2$ with the helicities $\lambda_1$ and
$\lambda_2$ are given by the
expressions
\beqa
\A_{1,1}^{(0)} & = &  \A_{-1,-1}^{(0)}=
  i\frac{2^{13}}{9\sqrt{3}}\pi^3\alpha_s^2\epsilon^2
  \frac{|R'(0)|}{M^{4}}f_T^2I_{1,1}^{(0)}(\epsilon), \label{eq:A11}\\ %
\A_{0,0}^{(0)} & = &
  -i\frac{2^{12}}{9\sqrt{3}}\pi^3\alpha_s^2
  \frac{|R'(0)|}{M}f_L^2I_{0,0}^{(0)}(\epsilon), \label{eq:A00}
\eeqa
where
\beqa
  \epsilon & = & m/M,
\eeqa
$m$ and $M$ are masses of vector and scalar mesons respectively, $R(r)$ is the radial part of the
scalar meson's wave
function, $f_{L,T}$ are longitudinal and transverse leptonic constants of the vector meson and
coefficients
$I^{(0)}_{\lambda_1,\lambda_2}$ are equal to
\beqa
  I_{1,1}^{(0)} & = &
    -\frac{1}{32}\int\limits _{0}^{1}dx\, dy \phi_T(x)\phi_T(y)
    \frac{1}{xy+(x-y)^2\epsilon^2}
    \frac{1}{(1-x)(1-y)+(x-y)^2\epsilon^2} \times
\nonumber \\ & \times &
   \frac{1}{2xy-x-y+2(x-y)^2\epsilon^2}
   \left[
     1+\frac{1}{2}\frac{(x-y)^2(1-4\epsilon^2)}
     {2xy-x-y+2(x-y)^2\epsilon^2}
   \right],\label{eq:I11}
\eeqa
\beqa
  I_{0,0}^{(0)} & = &
    -\frac{1}{32}\int\limits _0^1 dx\, dy \phi_T(x)\phi_T(y)
    \frac{1}{xy+(x-y)^2\epsilon^2}
    \frac{1}{1-x)(1-y)+(x-y)^2\epsilon^2} \times
\\  & \times &
    \frac{1}{2xy-x-y+2(x-y)^2\epsilon^2}
    \left\{
      1-\frac{1}{2}\frac{(x-y)^2(1-4\epsilon^2)}{2xy-x-y+2(x-y)^2\epsilon^2}-
\right.\nonumber \\  & - & \left.
    -2\epsilon^{2}\left[
       1+\frac{1}{2}\frac{(x-y)^2(1-4\epsilon^2)}{2xy-x-y+2(x-y)^2\epsilon^2}
     \right]\right\} .\label{eq:I00}
\eeqa
In the above equations $x$ and $y$ are the momentum fractions of the final mesons, carried by
quarks and
$\phi_{L,T}(x)$ are longitudinal and transverse distribution functions of these quarks in mesons.
Later we will
show, that the contribution of longitudinally polarized mesons to this process is small and one
can safely neglect
it, as it was done in \cite{Kartvelishvili:1984en}.

In \cite{Anselmino:1992rw} the similar formulae for nonzero helicity amplitudes of tensor meson
decay are also presented:
\beqa
  \A^{(2)}_{\lambda_1\lambda_2;\mu} & = &
    \tilde\A_{\lambda_1\lambda_2}e^{i\mu\varphi}d^{(2)}_{m,\lambda_1-\lambda_1}(\theta),
\eeqa
where $\mu$ is the meson spin projection on fixed axe, $\theta$ and $\varphi$ are polar and
azimuthal angles of one
of the final mesons in $\chi_2$ rest frame and reduced amplitudes $\tilde\A_{\lambda_1\lambda_2}$
are given by the expressions
\beqa
  \tilde\A_{1,1} & = & \tilde\A_{-1,-1} =
    -i \frac{2^{13}\sqrt{2}}{9\sqrt{3}}\pi^3\alpha_s^2\epsilon^2\frac{|R'(0)|}{M^4}f_T^2 I^{(2)}_
    {1,1} ,
  \\
  \tilde\A_{1,0} & = & \tilde\A_{0,1}=\tilde\A_{-1,0}=\tilde\A_{0,-1} =
    -i \frac{2^{12}\sqrt{2}}{9}\pi^3\alpha_s^2\epsilon\frac{|R'(0)|}{M^4}f_T f_L I^{(2)}_{1,0} ,
  \\
  \tilde\A_{1,-1} & = & \tilde\A_{-1,1} =
    -i \frac{2^{12}}{9}\pi^3\alpha_s^2\frac{|R'(0)|}{M^4}f_T^2 I^{(2)}_{1,-1} ,
  \\
  \tilde\A_{0,0} & = &
    -i \frac{2^{11}\sqrt{2}}{9\sqrt{3}}\pi^3\alpha_s^2\frac{|R'(0)|}{M^4}f_L^2 I^{(2)}_{0,0} ,
\eeqa
\beqa
  I^{(2)}_{1,1} & = & I^{(0)}_{1,1},
\\
  I^{(2)}_{1,0} & = &
    -\frac{1}{32}\int\limits_0^1\int\limits_0^1 dx dy \phi_T(x)\phi_L(y)
    \frac{1}{xy+(x-y)^2\epsilon^2} \frac{1}{(1-x)(1-y)+(x-y)^2\epsilon^2}
  \times \\ & \times &
    \frac{1}{2xy-x-y+2(x-y)^2\epsilon^2} \left[
      1+\frac{1}{2}\frac{(x-y)^2(1-4\epsilon^2)}{2xy-x-y+2(x-y)^2\epsilon^2}
    \right] ,
\eeqa
\beqa
  I^{(2)}_{1,-1} & = &
    -\frac{1}{32}\int\limits_0^1\int\limits_0^1 dx dy \phi_T(x)\phi_L(y)
    \frac{1}{xy+(x-y)^2\epsilon^2} \frac{1}{(1-x)(1-y)+(x-y)^2\epsilon^2}
  \times \\ & \times &
    \frac{1}{2xy-x-y+2(x-y)^2\epsilon^2},
\eeqa
\beqa
  I^{(2)}_{0,0} & = &
    -\frac{1}{32}\int\limits_0^1\int\limits_0^1 dx dy \phi_T(x)\phi_L(y)
    \frac{1}{xy+(x-y)^2\epsilon^2} \frac{1}{(1-x)(1-y)+(x-y)^2\epsilon^2}
  \times \\ & \times &
  \frac{1}{2xy-x-y+2(x-y)^2\epsilon^2} \left\{
    1+\frac{(x-y)^2(1-4\epsilon^2)}{2xy-x-y+2(x-y)^2\epsilon^2}+ \right.
  \\ &+& \left.
    4\epsilon^2\left[
      1+\frac{1}{2}\frac{(x-y)^2(1-4\epsilon^2)}{2xy-x-y+2(x-y)^2\epsilon^2}
    \right] \right\}.
\eeqa

In the literature plenty of different expressions for distribution functions $\phi_{L,T}(x)$ can
be found (review is
given in \cite{Chernyak:1983ej}) and, as it was shown in \cite{Ma:2004qf}, the result depends
strongly on the their
choice. That is why in our calculations we will use several variants:
\begin{itemize}
\item
Chernyak, Zhitnitsky \cite{Anselmino:1992rw} (in the tables this set will be labeled as "CZ") :
\beqa
  \phi_{L}\left(x\right) & = & \phi_{\textrm{CZ}}\left(x\right),\qquad\phi_{L}\left(x\right)=
  \phi_{2}\left(x\right),
\eeqa

\item Assymptotic \cite{Anselmino:1992rw} ("$\phi_1$") :
\beq
  \phi_{L}\left(x\right) & = & \phi_{T}\left(x\right)=\phi_{1}\left(x\right),
  \label{eq:Assymp}
\eeq

\item Symmetric \cite{Kartvelishvili:1977pi} ("$\phi_3$")
\beqa
  \phi_{L}\left(x\right) & = & \phi_{T}\left(x\right)=\phi_{3}\left(x\right),
\eeqa

\item $\delta$-approximation ("$\delta$")
\beqa
\phi_{L}\left(x\right) & = & \phi_{T}\left(x\right)=\delta\left(x-\frac{1}{2}\right).
\eeqa
\end{itemize}
%@@
All these distributions are normalized by the condition
\beqa
  \int\limits _{0}^{1}\phi\left(x\right)dx & = & 1
\eeqa
and the effective width of the distribution $\phi$ is defined as
\beqa
  \left\langle \delta x\right\rangle _{\phi} & = &
    \int\limits _{0}^{1}dx\phi^{2}\left(x\right)\left(x-\frac{1}{2}\right)^{2}.
\eeqa
In the above equations
\beqa
  \phi_{{\rm CZ}}\left(x\right)=13.2x\left(1-x\right)-36x^{2}\left(1-x\right)^{2},
& \qquad &
  \left\langle \delta x\right\rangle _{\phi_{{\rm CZ}}}\approx 6.7\cdot10^{-2},
\\
  \phi_{1}\left(x\right)=6x(1-x),
& \qquad &
  \left\langle \delta x\right\rangle _{\phi_{1}}\approx 5\cdot10^{-2},
\\
  \phi_{2}\left(x\right)=30x^2(1-x)^2,
& \qquad &
  \left\langle \delta x\right\rangle _{\phi_{1}}\approx 3.6\cdot10^{-2},
\\
  \phi_{3}\left(x\right)=140x^{3}\left(1-x\right)^{3},
& \qquad &
  \left\langle \delta x\right\rangle _{\phi_{3}}\approx 2.8\cdot10^{-2}.
\eeqa
By analogy with the reaction $\epem\to\JP\ec$ one can expect, the branching fraction of the decay
$\chi_{b0,2}\to\JP\JP$
will increase with the increase of the width of the distribution.

In our work we consider the decays $\chi_{c0,2}\to\rho\rho$, $\chi_{c0,2}\to\phi\phi$ and
$\chi_{b0,2}\to\JP\JP$.
Numerical values for parameters are listed in table \ref{tab:dec}. The values of branching
fractions and decay widths
obtained with the help of these distributions are presented in table \ref{tab:GBr}. From this
table one can easily see that
the branching fractions do strongly depend on the distribution functions and increase with
increase of their "widths". As
it was mentioned above, the experimental value of the $\chi_{c0}\to\phi\phi$ branching fraction
is much greater than the
theoretical predictions made in the framework of $\delta$-approximation.
Now internal motion of quarks in mesons raises these predictions. For example, with the
distribution (\ref{eq:Assymp}) we get
\beqa
\Br(\chi_{c0}\to\phi\phi) & = & 0.9\cdot 10^{-3},
\eeqa
while the experimental result is $(1.0\pm0.6)\cdot10^{-3}$. The same can be said about the decays
$\chi_{b0,2}\to\JP\JP$.
The calculations held in $\delta$-approximation give $\Br(\chi_{b0}\to\JP\JP)=1.4\cdot10^{-5}$,
and now this value can be
raised to $2\cdot10^{-4}$.

In table \ref{tab:Ratio} we present the ratio of helicity amplitudes for production of
longitudinally and transversely
polarized vector mesons. From this table it is clear that this ratio is less than $15\%$ for all
the considered decays and
the production of longitudinally polarized mesons can be safely neglected, as it was done in
\cite{Kartvelishvili:1984en}.

%%%%%%%%%%%%%%%%%%%%%%%%%%%%%%%%%%%%%%%%%%%%%%%%%%%%
\section{$p\bar p \to \JP\JP X$}

As it was mentioned in \cite{Kartvelishvili:1984en}, the decay $\chi_{b0,2}\to\JP\JP$ can be
utilized in experiment to
study the properties of $\chi{bJ}$ mesons in high energy reactions (for example, in the inclusive
process
$p\overline{p}\to ggX\to\chi_{b0,2}X\to\JP\JP X$). It should be mentioned, that in addition to
resonance production of
$\JP\JP$ pair via $\chi_{bJ}$ decays there is also a non-resonance production of $\JP\JP$ pair in
continuum and it is
important to study the possibility of separating of the resonance signal from this background.

The cross section of the double $\JP$ production in the proton-antiproton annihilation can be
related to the cross section
of its partonic subprocess $\hat\sigma(ab\to\JP\JP)$ and the structure functions of the initial
hadrons according to the
relation
\beq
  \sigma(p\bar p\to abX \to \JP\JP X) & = &
  \sum\limits_{a,b} \int d\xi_a d\xi_b f^{(p)}_a(\xi_a) f^{(\bar p)}_b(\xi_b)\hat\sigma(ab\to\JP
  \JP)
\label{eq:hadr},
\eeq
where $\xi_a$ and $\xi_b$ are the momentum fractions of partons $a$ and $b$, $f^{(p)}_a(\xi_a)$
and $f^{(\bar p)}_b(\xi_b)$
are the distribution functions of partons $a$ and $b$ in proton and antiproton. It is convenient
to use new integration
variables $\hat s\approx s\xi_a\xi_b$ --- the square of the invariant mass of the partonic pair
and $\xi=\xi_a-\xi_b$. If
we are interested in the central production of $\JP\JP$ pair in the resonance region, we have
$\xi_a=\xi_b=M_\chi/\sqrt{s}\ll 1$. In this case the gluonic pair (i.e. $a=b=g$) gives the main
contribution to the
reaction
give and the equation (\ref{eq:hadr}) can be written in the form
\beq
  \frac{d\sigma(p\bar p\to\JP\JP X)}{d\hat s} & = &\frac{1}{16} \frac{\hat\sigma_{gg}}{\hat s} L,
  \label{eq:hadr2}
\eeq
where the first factor corresponds to averaging over gluon colors and helicities,
$\hat\sigma_{gg}=\hat\sigma(gg\to\JP\JP)$, and the dimensionless factor
\beq
  L(\hat s) &=& 2\int\limits_0^{1-\hat s/s}\frac{\xi_a\xi_b}{\xi_a+\xi_b} f^{(p)}_g(\xi_a) f^{(
  \bar p)}_g(\xi_b) d\xi
  \label{eq:L}
\eeq
describes the partonic luminosity.

The partonic cross section $\hat\sigma_{gg}$ equals the sum of cross sections of resonance
reaction
$p\bar p\to ggX\to \chi_{bJ}X\to\JP\JP X$ and the non-resonance background:
\beqa
  \hat\sigma_{gg} & = & \hat\sigma_r + \hat\sigma_{nr}
\eeqa

Resonance part of the cross section can be obtained from the values presented in table
\ref{tab:GBr} with the help of
Breit-Wigner formula:
\beq
  \hat{\sigma}_{{\rm r}} & = &
    \sum\limits_{J=0,2} \frac{\pi}{M_{gg}^2} (2J+1)
    \frac{\Gamma(\chi_{bJ}\to gg)\Gamma(\chi_{bJ}\to\JP\JP)}{(M_{gg}-M_{\chi_{bJ}})^2+\Gamma_{
    \chi_{bJ}}^2/4},
    \label{eq:sr}
\eeq
where  $M_{gg}=\sqrt{\hat{s}}$ --- invariant
mass of the colliding gluons and total widths of $\chi_{bJ}$-mesons can be obtained from the
values of the radial part
$R(r)$ of their wave functions in the origin \cite{Olsson:1984im}:
\beqa
  \Gamma_{\chi_{b0}} & = &
    \Gamma(\chi_{b0}\to gg) = 96\frac{\alpha_s^2}{M_{\chi_{b0}}^4} |R'(0)|^2 \approx 0.544\MeV,
    \\
  \Gamma_{\chi_{b2}} & = &
    \Gamma(\chi_{b2}\to gg) = \frac{128}{5}\frac{\alpha_s^2}{M_{\chi_{b2}}^4} |R'(0)|^2 \approx
    0.14\MeV.\\
\eeqa

The implementation error $\Delta$ could however be much smaller than this widths and it can be
useful to take it into
account. This can be done by means of simple substitution
\beq
  \hat\sigma_r^\Delta & = &
    \sum\limits_{J=0,2} \frac{\pi}{M_{gg}^2} (2J+1)\frac{\Delta}{\Gamma_{\chi_{bJ}}}
    \frac{\Gamma(\chi_{bJ}\to
    gg)\Gamma(\chi_{bJ}\to\JP\JP)}{(M_{gg}-M_{\chi_{bJ}})^2+\Delta^2/4}.
    \label{eq:srD}
\eeq
It can be easily seen that this substitution does not change the value of the total integrated
cross section (i.e.
$\int\hat\sigma_r dM_{gg}=\int\hat\sigma_r^\Delta dM_{gg}$).

Non-resonance part of the cross section (i.e. cross section of the double $\JP$ production in
continuum) was obtained in
\cite{Humpert:1983qt,Kartvelishvili:1983aa}. According to this works
\beq
  \frac{d\hat\sigma_{nr}}{d\hat t} & = &
    \frac{3}{4\pi}\left(
      \frac{\pi\alpha_s}{3}\right)^4 \left(\frac{|\psi(0)|^2}{M_\psi}
    \right)^2 \frac{\hat s^4 \sum\limits_{n=0}^4y^n C_n}{(\hat t-M_\psi^2)^4(\hat u-M_\psi^2)^4},
    \label{eq:snr}
\eeq
where $\hat s$, $\hat t$ and $\hat u$ are the Mandelstam variables of the partonic subprocess,
$M_\psi$ and $\psi(0)$ are mass and wave function of $\JP$, $y=(\hat t+\hat u)/2\hat s$,
$z=(\hat t-\hat u)/2\hat s$,
and the coefficients $C_i$ equals
\beqa
C_0 & = & \frac{2}{3}\left(335 - 13568\,z^2 + 98144\,z^4 + 35584\,z^6 - 1186048\,z^8 + 258048\,z^
{10} + 3981312\,z^{12}
\right) ,\\
C_1 & = & \frac{8}{3}\left(137 - 14288\,z^2 + 168064\,z^4 - 403840\,z^6 - 596224\,z^8 + 1935360
\,z^{10}\right) ,\\
C_2 & = & \frac{8}{3}\left(-1 - 11472\,z^2 + 239008\,z^4 - 1010944\,z^6 + 1134336\,z^8\right) ,\\
C_3 & = & \frac{64}{3}\left(5 + 828\,z^2 + 688\,z^4 - 10176\,z^6\right) ,\\
C_4 & = & 256\left(1 + 8\,z^2 + 16\,z^4\right).
\eeqa

\begin{figure}
  \begin{picture}(400,210)
  \put(0,0){\epsfxsize=15cm \epsfbox{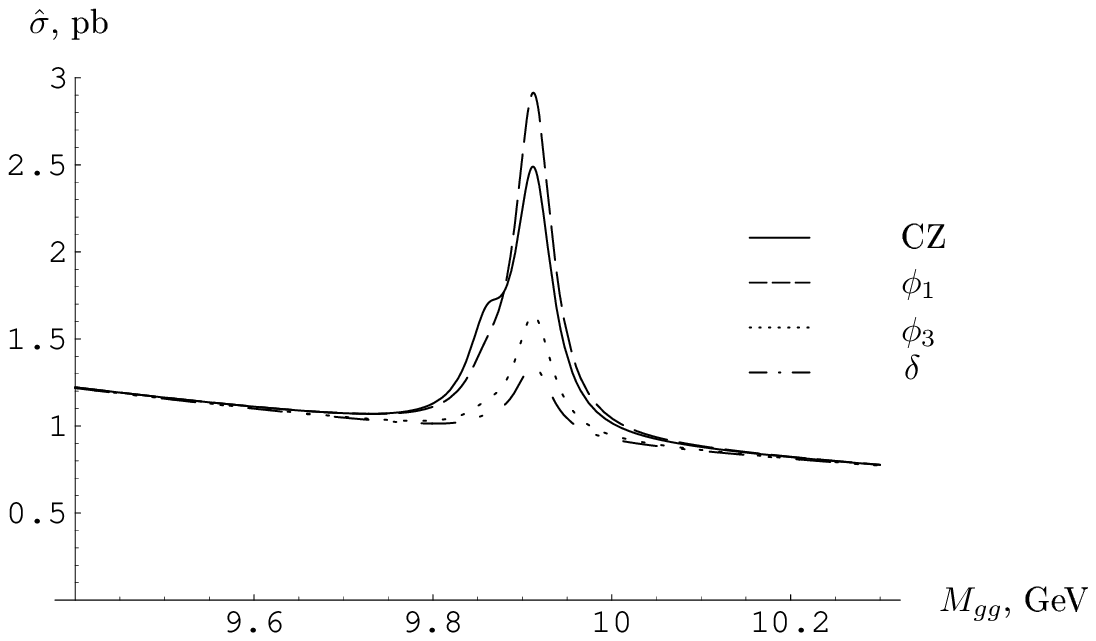}}
  \end{picture}
  \caption{Distribution of the partonic cross section versus gluon-gluon mass}\label{fig:dsdMgg}
\end{figure}

\begin{figure}
  \begin{picture}(400,300)
  \put(0,0){\epsfxsize=15cm \epsfbox{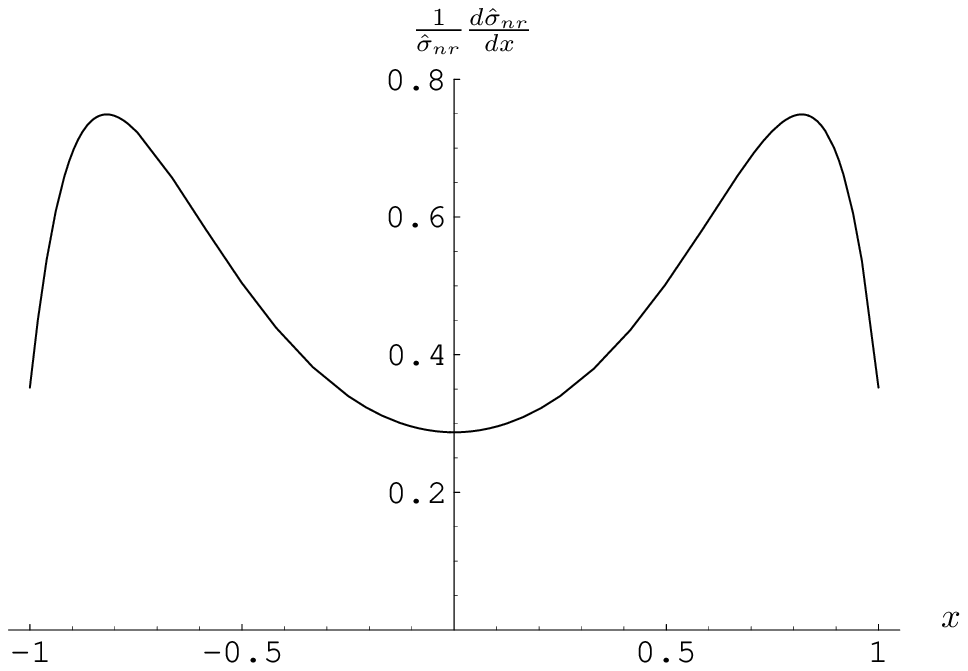}}
  \end{picture}
  \caption{Angular distribution for the background process}\label{fig:DsDx}
\end{figure}

\begin{figure}
  \begin{picture}(400,200)
  \put(0,0){\epsfxsize=15cm \epsfbox{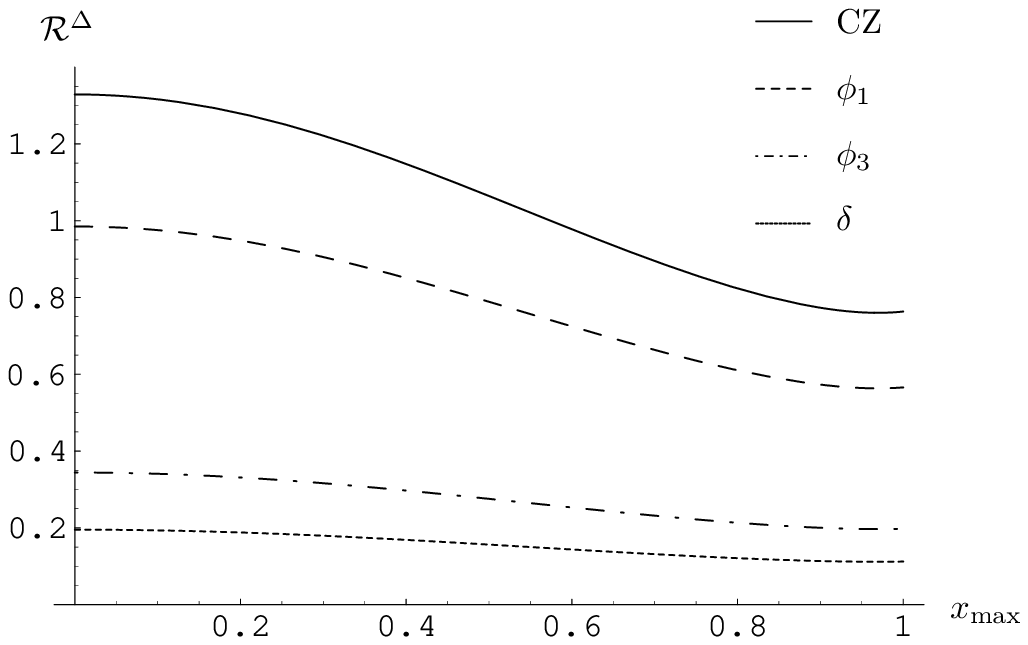}}
  \end{picture}
  \caption{Signal/background ratio for different distribution functions}\label{fig:sigmaXcut}
\end{figure}

On fig.\ref{fig:dsdMgg} the dependence of partonic cross section form the invariant mass of
colliding gluons in the
resonance mass region and the value for the implementation error $\Delta=50$ MeV is shown. It is
clearly seen from this
figure that the separation of $\chi_b$ signal from the background for most of distributions
should not make any troubles.
On the other hand for the $\delta$-approximation the signal/background ratio
\beqa
  \R^\Delta & = & \left. \frac{\hat\sigma_r^\Delta}{\hat\sigma_{nr}}\right|_{M_{gg}=M_\chi}
\eeqa
is not high enough and it may be useful to study the possibilities of increasing this ratio. It
can be made by applying
certain cuts on the kinematics of final $\JP$ mesons. As it can bee seen in fig.\ref{fig:DsDx},
where the angular
distribution of the background cross section over the cosine of scattering angle $x=\cos\theta$
is shown, final $\JP$
mesons fly mainly along the beam axes, while the distribution of $\JP$ produced in $\chi_b$
decays is isotropic. Thats why
applying the condition $\left|x\right|<x_{max}$ we can increase the ratio $\R^\Delta$. On fig,
\ref{fig:sigmaXcut} the
dependence of this ratio form the cut angle is shown.

Let us now return to the total cross section of the $\chi_{b0,2}$-meson production. According to
formula (\ref{eq:hadr2})
this cross section is equal to
\beq
\sum\limits_{J=0,2}\sigma(p\bar p\to \chi_{bJ}+X) & = &
 \frac{1}{16} \int \frac{d\hat s}{\hat s} L(\hat s) \hat\sigma_r(\hat s),
  \label{eq:sigma}
\eeq
where $\hat\sigma_r(\hat s)$ is the resonance cross section (\ref{eq:sr}). Because the resonances
are narrow, we can
neglect the dependence of the partonic luminosity $L(\hat s)$ on $\hat s$, so the integral over
$\hat s$ in
(\ref{eq:sigma}) will reduce to a simple expression:
\beqa
\sum\limits_{J=0,2}\sigma(p\bar p\to \chi_{bJ}+X) & = &
 \frac{\pi^2}{4M_\chi^3} L \sum\limits_{J=0,2} (2J+1)\Gamma(\chi_{bJ}\to gg),
\eeqa
where the value of the partonic luminosity $L$ is defined according to formula (\ref{eq:L}). For
the gluon distribution
functions in the initial (anti)protons we have used the parameterization presented in \cite{
Alekhin:2002fv} at the value of
virtuality $Q^2=M_\chi^2$. The values of $\chi_{b0}$ and $\chi_{b2}$ production cross sections
obtained using this formula
are equal to 
\beqa
\sigma(p\bar p\to \chi_{b0}+X)=250\nb,\qquad\sigma(p\bar p\to\chi_{b2}+X)=320\nb 
\eeqa
at the Tevatron energies ($\sqrt{s}=2$ TeV) and 
\beqa
\sigma(p\bar p\to\chi_{b0}+X)=1.5\mcb,\qquad \sigma(p\bar p\to\chi_{b2}+X)=2\mcb
\eeqa
at LHC energies ($\sqrt{s}=14$ TeV). In the $\JP\JP$ mode these values correspond to 
\beqa
\sigma(p\bar p\to\chi_{b0}+X)\Br(\chi_{b0}\to\JP\JP)=53\pb,\qquad
\sigma(p\bar p\to\chi_{b2}+X)\Br(\chi_{b2}\to\JP\JP)=170\pb
\eeqa
at $\sqrt{s}=2$ TeV and
\beqa
\sigma(p\bar p\to\chi_{b0}+X)\Br(\chi_{b0}\to\JP\JP)=330\pb,\qquad
\sigma(p\bar p\to\chi_{b2}+X)\Br(\chi_{b2}\to\JP\JP)=1\nb
\eeqa
at $\sqrt{s}=14$ TeV. 

The possibility of quarkonia observation in their $\JP\JP$ decays was also proposed in the works
\cite{Braaten:2000cm,Maltoni:2004hv}. The value of the $\eta_b$ production cross section at Tevatron presented there
\beqa
\sigma(p\bar p\to\eta_b+X)=2.5\mcb
\eeqa
is about an order of magnitude higher, then our results. This difference can be explained by the difference in the
total widths of $\eta_b$ and $\chi_b$ mesons. Inspired by these works the CDF collaboraion has measured the $\JP\JP$ 
mass distribution  and has observed a small cluster in the $\eta_b$ region \cite{Tseng:2003md}
\footnote{
The authors are grateful to N. Brambilla for drawing our attention to this work.
}. 
Since this cluster is small only upper 95\% CL upper limit on 
$\sigma_{\eta_b}(|y|<0.4)\Br(\eta_b\to\JP\JP)[\Br(\JP\to\mu^+\mu^-)]^2$ was found in this experiment. As it can be seen
from the $\JP\JP$ mass spectrum, there are also some events in the $\chi_b$ region, so
one can hope that with the increase of luminocity it 
will become possible to observe $\chi_b$ mesons in the reaction studied in this paper.

It should be mentioned, that our estimates were
obtained in the collinear
gluon approximation, i.e. using the integrated over the transverse momentum gluon structure
functions. As a result of this
approximation we have underestimated the values of the $\chi_b$ meson production cross section
and it is comparable with
the cross section of direct $\Upsilon$ production \cite{Shears:2003ic}. So, our estimates should
be considered as lower
bounds and the situation could be more favorable for the $\chi_b$ observation in $\JP\JP$ mode.

\section{Conclusion}

The discrepancy of experimental results and theoretical predictions for the cross section of
$\JP\ec$ production in $\epem$
annihilation have found a natural explamation by taking into account intrinsic motion of quarks
in $\JP$ and $\ec$ mesons
in
the amplitude of heavy quarks production. At this point our results coincide with the results of
the works \cite{
Ma:2004qf,Bondar:2004sv}.

We have also studied another example of the reaction $\chi_{0,2}\to VV$, and specifically the
decays of $\chi_{b0}(0^{++})$
and $\chi_{b2}(2^{++})$	into a pair of $\JP$ mesons. It is shown, that taking into account the
relative motion of quarks in
the amplitude of the decay of $\chi_b$ meson to $c$-quarks increase the branching fractions of
these decays by an order of
magnitude. The agreement with the experiment was achieved for the only experimentally known decay
$\chi_{c0}\to\phi\phi$.

Later we have considered the possibility of the observation of $\chi_b\to\JP\JP$ decay in the
experiment. The values of the
cross sections of $\chi_b$ production in $p\bar p$ annihilation and its subsequent decay into
$\JP\JP$ pair presented in
our paper show, that it is possible to observe this reaction at Tevatron and LHC energies.

%%%%%%%%%%%%%%%%%%%%%%%%%%%%%%%%%%%%%%%%%%%%%%%%%
\begin{acknowledgments}
The authors thank professor S.S. Gershtein and D. Denisov for useful discussions. This work was partially
supported by Russian Foundation of Basic Research under grant 04-02-17530, Russian Education
Ministry grant E02-31-96, CRDF grant MO-011-0, Scientific School grant SS-1303.2003.2. One of
the authors (V.B.) was also supported by Dynasty foundation.

\end{acknowledgments}

%%%%%%%%%%%%%%%%%%%%%%%%%%%%%%%%%%%%%%%%%%%%%%%%%%%%%%%%%%%%%%%%%%%
\bibliographystyle{C:/texmf/bibtex/bst/revtex4/apsrev}

\clearpage\newpage

\begin{table}
\caption{Cross sections of the $\epem\to\JP\ec$ process using different parametrizations of meson wave functions}\label{tab:tab}
$$\begin{array}{|c|c|c|c|c|c|}
\hline
v^2  &  |R(0)|^2 ( \mbox{GeV} )  &  \sigma_0 (\mbox{fb}) & \sigma/ \sigma_0 & (\sigma/ \sigma_0)_
T & \sigma (\mbox{fb})
\\
\hline
0.2 & 0.31 & 0.78 & 1.9 & 2.3 & 1.7 \\
0.3 & 0.50 & 1.9 & 2.4 & 3.5 &  6.5\\
1/3 & 0.55 & 2.3 & 2.6 & 4.0 &  9.2 \\
0.4 & 0.65 & 3.2 & 2.9 & 5.2 & 16.6 \\
\hline
\end{array}$$
\end{table}

\begin{table}
  \caption{Parameters
  \label{tab:dec}}
$$\begin{array}{|c|c|c|c|c|c|c|}
\hline
 & M,\mbox{ÃýÂ} & m,\mbox{GeV} & \alpha_s & f_T,\mbox{GeV} & f_L,\mbox{GeV} &
 |R'(0)|^2,\mbox{GeV}^5 \\
\hline
\chi_{c0}\to\rho\rho & 3.415 & 0.776 & 0.3 & 0.2 & 0.2 & 0.11\\
\chi_{c0}\to\phi\phi & 3.415 & 1.02 & 0.3 & 0.2 & 0.2 & 0.11\\
\chi_{b0}\to\JP\JP & 9.8599 & 3.09692 & 0.2 & 0.4 & 0.4 & 1.34\\
\hline
\chi_{c2}\to\rho\rho & 3.55626 & 0.776 & 0.3 & 0.2 & 0.2 & 0.11\\
\chi_{c2}\to\phi\phi & 3.55626 & 1.02 & 0.3 & 0.2 & 0.2 & 0.11\\
\chi_{b2}\to\psi\psi & 9.9126 & 3.09692 & 0.2 & 0.4 & 0.4 & 1.34\\
\hline
\end{array}$$
\end{table}

\begin{table}
\caption{
$\G(\chi\to VV)$, keV / $\Br(\chi\to VV)$, $10^{-4}$
}
$$\begin{array}{|c|c|c|c|c|}
 \hline
 & \mathrm{CZ} & \phi_1 & \phi_3 & \delta\\
\hline
\chi_{c0}\to\rho\rho & 24.9\, /\, 35.6 & 8.98\, /\, 12.9 & 2.75\, /\, 3.93 & 1.36\, /\, 1.94\\
\chi_{c0}\to\phi\phi & 16.3\, /\, 23.3 & 6.31\, /\, 9.02 & 2.10\, /\, 3.01 & 1.09\, /\, 1.57\\
\chi_{b0}\to\JP\JP & 0.117\, /\, 2.15 & 0.0462\, /\, 0.849 & 0.0157\, /\, 0.288 & 0.00827\, /\,
0.152\\
\hline
\chi_{c2}\to\rho\rho & 6.94\, /\, 43.8 & 10.6\, /\, 66.6 & 3.39\, /\, 21.4 & 1.89\, /\, 12.\\
\chi_{c2}\to\phi\phi & 7.14\, /\, 45.1 & 9.79\, /\, 61.8 & 3.38\, /\, 21.3 & 1.96\, /\, 12.4\\
\chi_{b2}\to\psi\psi & 0.0757\, /\, 5.32 & 0.100\, /\, 7.05 & 0.0355\, /\, 2.5 & 0.0208\, /\,
1.47\\
\hline
\end{array}$$
\label{tab:GBr}
\end{table}

\begin{table}
\caption{
$2|\A^{0}_{1,1}/\A^{(0)}_{0,0}|^2$,\%
}
$$\begin{array}{|c|c|c|c|c|}
 \hline
 & \mathrm{CZ} & \phi_1 & \phi_3 & \delta\\
\hline
\chi_{c0}\to\rho\rho & 0.306 & 1.99 & 2.18 & 2.65\\
\chi_{c0}\to\phi\phi & 1.23 & 7.41 & 7.95 & 9.43\\
\chi_{b0}\to\JP\JP & 1.61 & 9.59 & 10.3 & 12.1\\
\hline
\end{array}$$
  \label{tab:Ratio}
\end{table}

\end{document}